\documentclass[12pt]{iopart}

\usepackage{dcolumn}% Align table columns on decimal point
\usepackage{bm}% bold math
\usepackage{graphicx}% Include figure files
\usepackage{color}
\usepackage{subfigure}
\usepackage{hyperref}
\usepackage{latexsym}
\usepackage{amsthm}
\usepackage{amssymb}
\usepackage{array}
\usepackage{braket}

\DeclareGraphicsExtensions{.jpg,.pdf, .mps, .png, .eps, .ps, .EPS,.gif}

\def\be{\\begin{equation}}
\def\ee{\end{equation}}

\def\bc{\begin{center}}
\def\ec{\end{center}}
\def\bea{\begin{eqnarray}}
\def\eea{\end{eqnarray}}
\newcommand{\avg}[1]{\langle{#1}\rangle}
\newcommand{\Avg}[1]{\left\langle{#1}\right\rangle}
\begin{document}

\title[Critical time-dependent branching process ]{Critical time-dependent branching process modelling epidemic spreading with containment measures}

\author{Hanlin Sun}

\address{School of Mathematical Sciences, Queen Mary University of London, London, E1 4NS, United Kingdom}

\author{Ivan Kryven}
\address{Mathematical Institute, Utrecht University, PO Box 80010, 3508TA Utrecht, the Netherlands}

\author{Ginestra Bianconi}
\address{ 
School of Mathematical Sciences, Queen Mary University of London, London, E1 4NS, United Kingdom\\
The Alan Turing Institute, The British Library, United Kingdom}

%\ead{submissions@iop.org}
\vspace{10pt}
\begin{indented}
\item[]June 2021
\end{indented}

\begin{abstract}
During the COVID pandemic, periods of exponential  growth of the disease have been mitigated by containment measures that in different occasions have resulted in a power-law growth of the number of cases. The first observation of such behaviour has been obtained from 2020 late spring data coming from China by Ziff and Ziff in Ref.\cite{ziff2020fractal}. After this important observation the power-law scaling (albeit with different exponents) has also been observed in other countries during periods of containment of the spread. Early interpretations of these results suggest  that this phenomenon might be due to spatial effects of the spread. Here we show that temporal modulations of infectivity of individuals due to  containment measures can also cause power-law growth of the number of cases over time.  
To this end we propose a stochastic well-mixed Susceptible-Infected-Removed (SIR) model of epidemic spreading  in presence of containment measures resulting in a time dependent infectivity and we explore the statistical properties of the resulting branching process at criticality. 
We show that at criticality it is possible to observe power-law growth of the number of cases with exponents ranging between one and two. Our asymptotic analytical results are confirmed by extensive Monte Carlo simulations. Although these results do not exclude that spatial effects might be important in modulating the power-law growth of the number of cases at criticality, this work shows that even well-mixed populations may already feature non trivial power-law exponents at criticality.
\end{abstract}

\noindent{\it Keywords}: Branching process, epidemic spreading, critical phenomena\\
%\maketitle

{\it This work celebrates the $70^{\mbox{th}}$ birthday of our dear friend and colleague Bob Ziff.}\\

\section{Introduction}

Exponential growth of the number of cases is typically observed at the onset of an epidemic when the dynamics is in the supercritical regime. The COVID data has also supported this claim and at the beginning of the current pandemic the scientific community has extensively confirmed exponential growth of the number of cases in different countries.
However Ziff and Ziff in Ref. \cite{ziff2020fractal}  were the first to detect a power-law growth in the number of cases  starting from data coming from the late spring of 2020 in China when the epidemic was suppressed by containment measures.
Later on the power-law growth of the number of cases has been recorded in data coming from other countries \cite{nekovee2020understanding,brandenburg2020piecewise}. Interestingly these results have been obtained in cases of successful containment of the epidemic spreading after the implementation of  efficient containment measures \cite{maier2020effective}, such as contact tracing (automatic and not), social distancing, testing and or other policies aimed at isolating timely infectious individuals and at reducing their reproductive number.

An important question that arises is: what is the mechanism responsible for the power-law scaling of the number of cases? Is this a phenomenon caused by the spatial distribution of the cases?   Is it the sign that the system is reaching a critical behaviour consistent with a $R_0=1$? Or can it be a combination of the latter two effects? If not, is this the effect of the containment measures?

During the current pandemic there has been a surge in research on epidemic spreading. Many works have discussed the challenges of epidemic spreading modelling \cite{carletti2020covid,vespignani2020modelling}, a number of works have addressed outstanding theoretical problems that the current pandemic has highlighted  \cite{bianconi2020epidemics,radicchi2020epidemic,krapivsky2021immortal,krapivsky2021infection,sun2021competition,st2021universal,vazquez2020superspreaders} and a vast attention has been devoted to  extract information from epidemic data  \cite{maier2020effective,pepe2020covid,bianconi2020efficiency,paul2021socio}. Additionally scientific research has informed  policy makers \cite{colizza_app,colizza_lock} establishing the role that containment measures   such as  social distancing,  or contact tracing   \cite{ferretti2020quantifying,bell2021beyond,bianconi2021message,cencetti2021digital,caccioli2020epidemic,
kryven2021contact} have in mitigating the epidemic spread. 
{Among the theoretical results we mention possible explanation of  the power-law scaling have been proposed including  interpretation of the power-law scaling as a signature of criticality \cite{bianconi2020epidemics,radicchi2020epidemic}, as an effect of the inhomogeneous network of contact \cite{vazquez2020superspreaders} or as due to the  fractal spatial  distribution of the spread \cite{ziff2020fractal}.}

Here we consider a very stylized theoretical model in a  well-mixed population that is simple enough to be analytically solvable neglecting many detailed aspects of the realistic epidemic spreading model, yet capturing important statistical aspects that go beyond the simplest branching process. 
We show that a power-law growth of the number of cases can be observed when the epidemic process reaches criticality due to containment measures that allow for a temporal modulation of the infectivity of infectious individuals.
 In particular, while the Susceptible-Infected-Removed model at criticality predicts a power-law growth of the number of infected individuals with a power-law exponent equal to two, here we show that containment measures can be responsible for modulating the power-law exponent between one and two.
In order to demonstrate this modulation of the dynamical critical exponent we propose a discrete time epidemic model based on a branching process in which an infected seed individual can infect a different number of individuals at each time during seed's infectious period.
This branching process is characterized by the distribution $D(t)$ of the duration of the infectious period of each infected individual and the function $m(t')$ indicating the expected number of individuals infected by an infectious individual after time $t'$ from contracting an infection.
This model is chosen to capture a temporal modulation of the infectivity of the infectious individual and clearly differs from the age-dependent branching process 
\cite{grimmett,age1,age2} where each infected individual gives rise to new infected individuals  at a single time, even if this time is chosen randomly.
We characterise the critical properties of the proposed branching process as a function of $D(t)$ and $m(t)$, derive the critical indices of the dynamics and compare the results with extensive Monte Carlo simulations.  As expected, this analysis reveals that stochastic effects play a key role in determining these exponents, which may strongly deviate from the exponents in deterministic approaches \cite{bianconi2020epidemics}.
Moreover, these results show that time-dependent modulation of the infectivity can be responsible for a modulation of the power-law exponent determining the power-law growth of the number of cases in time.
We note that these results do not exclude a priori that spatial effects might also be important elements determining the power-law increase in the number of cases. In particular, hierarchical and hyperbolic networks describing nested communities of people during lockdown can be responsible for a broadening of the critical region in which one can observe the power-law critical behaviour \cite{moretti2013griffiths} similarly to what happens for percolation on the same type of networks     \cite{boettcher2012ordinary,bianconi2018topological,kryven2019renormalization,bianconi2019percolation,sun2020renormalization}.

\section{Epidemic spreading with containment measures}
\subsection{The major properties of the SIR model}
The Susceptible-Infected-Removed (SIR) model is a well-known model of epidemic spreading in which individuals can be in one of three possible states: 
1) susceptible can get infected when in contact with an infectious individual, 
2) infected can spread the infection to susceptible individual upon contact with infectivity rate $\lambda$, and 
3) removed or recovered cannot spread the infection anymore. 
 This model is known to display three dynamical regimes depending on the value of infectivity: for $\lambda>\lambda_c$ the epidemics is in the supercritical regime, when the epidemic affects a positive fraction of the population; b) for $\lambda<\lambda_c$  the subcritical regime is observed, when the epidemic dies out before spreading in the population,  and c) for $\lambda=\lambda_c$  the epidemics is in the critical regime, when the epidemics affects a sublinear fraction of all individuals. Here, $\lambda_c$ indicates the so-called {\em epidemic threshold}. 
 However, it has to be noted that in hyperbolic and hierarchical structures the critical region  may stretch out for a finite range of values of the infectivity \cite{moretti2013griffiths}, which corresponds to the fact that in these networks one can observe two percolation thresholds   \cite{boettcher2012ordinary,bianconi2018topological,kryven2019renormalization,bianconi2019percolation,sun2020renormalization}.
When the onset of the epidemic is started from a single infected individual, the latter three dynamical regions are characterised by different dynamical properties: the supercritical region is characterised by an exponential increase of the number of infected individuals, the critical regime -- by a power-law increase with exponent $2$, while the subcritical regime -- by finite size stochastic fluctuations.

\subsection{Introducing a time dependent infectivity}

In a typical Susceptible-Infected-Removed  (SIR) epidemic model it is assumed that infectivity $\lambda$ does not change with time as long as an infected individual is contagious. In other words, the total number of secondary infections is proportional to the time an individual was infectious. Moreover, it is assumed that each infected individual is removed from the population with a probability that does not depend on time.

Here we consider a model in which each infected individual has a reproductive number that depends on the time elapsed since his/her  infection. Hence we consider time-dependent infectivity by substituting
\begin{equation}
\lambda\to \lambda F(t),
\end{equation}  
 where $F(t)$ is a decreasing function of $t$, indicating the time elapsed since the infection of the infectious individual.
We additionally assume that the probability that an infectious individual recovers is also time-dependent.
This model can be considered as the stochastic model underlying the deterministic dynamics proposed in Ref. \cite{bianconi2020epidemics}.
The decay of the effective infectivity can be due to different causes, including  asymptomatic onset,  early testing policies, and containment measures enforced once the infection becomes symptomatic, {\it i.e.}  the transmission time.
In the supercritical regime this model can be treated using a deterministic approach, which predicts an exponential increase in the number of infected individuals at the onset of the epidemics. In order to perform the asymptotic analysis of this process we consider the scenario of an infinite population.

Our model has discrete time. By taking the moment when an individual becomes infections as a reference, we denote the time that has elapsed since this event as $t=1,2,\dots$
We then assume that at every time step $t>0$, this individual recovers/ is removed with probability $q(t)=1-p(t)$ or remains to be infectious with probability $p(t)$. Therefore the probability that the infected individual is still infectious at $t$ is given by 
\begin{equation}
P(t)=\prod_{t^{\prime}=1}^{t}p(t^{\prime}).
\end{equation}
Additionally, we assume that at time $t$,  an individual transmits infection to  $z_{t}$  susceptible individuals. Here, $z_{t}$ is a random number drawn from the Poisson distribution with mean $\lambda m(t)$, where $m(t)$ is either a constant or a decreasing function of time. In expectation, an individual that recovers at time $t$ has a cumulative number of transmissions  given by
\bea
\lambda M(t)=\lambda\sum_{t^{\prime}=1}^{t-1} m(t),
\label{eq:Mt}
\eea
where we have assumed $m(0)=0$.\\
In this stochastic model it is immediate to show  that  an  infectious individual infects, in average,
\begin{equation}
\lambda F(t)=\lambda P(t) m(t)
\end{equation}
other individuals after time $t$.
It follows that  $F(t)$ acts as an overall dressing of the infectivity, capturing timely detection, tracking  and isolation of the cases. 
Let us indicate with $n(t)$ the average  number of newly infected individuals at timestep $t$.
Starting with a single individual infected at time $t=0$, {\it i.e.} $i(0)=1$, in average, the number of new infected individuals at time $t$ reads
\begin{equation}
{i}(t)=\lambda \sum_{t'=1}^{t-1} F(t-t')i(t'). \label{eq:new_infection}
\end{equation}
The expected number of  newly removed individuals $r(t)$ at time $t$ is then

\begin{equation}
r(t)=\sum_{t'=1}^{t-1}\prod_{t''=t'+1}^{t-1}p(t)[1-p(t''-t')]i(t').
\end{equation}

The average number $I(t)$ and $R(t)$ of infected and removed individuals at time $t$ is 
\bea
I(t)&=&\sum_{t'=1}^{t}P(t-t')i(t'), \\
R(t)&=&\sum_{t'=1}^{t}[1-P(t-t')]i(t').
\eea

\section{Time dependent branching process with containment measures}
\begin{figure*}
\begin{center}
	\includegraphics[width=0.8\columnwidth]{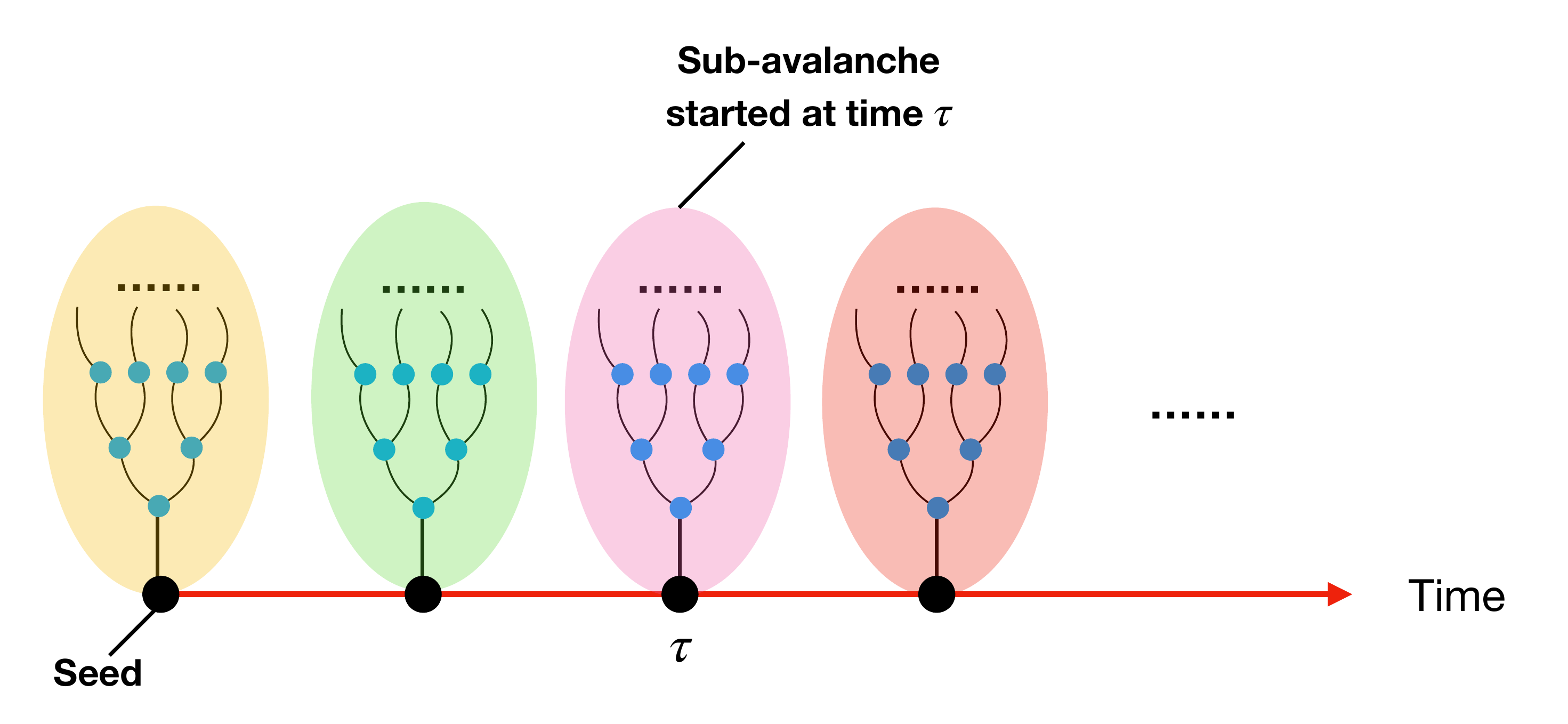} 
	\end{center}
	\caption{A schematic figure of the branching processes. The seed starts a new sub-avalanche at each time step during its infectious period. The avalanche size $n$ is given by summing up the size of all sub-avalanches and the seed itself. Note that each infected individual of any subavalanches will also produce a series of different subavalanches during each time step of its infectious period (not shown for simplifying the figure).
	 }\label{fig:branching_process}
\end{figure*}

The model described in a previous section can be studied by considering a branching process.
In this branching process the avalanche generated by a single node is due to the sum of subavalanches generated by each of the individuals infected by the seed node at any given time (see Figure \ref{fig:branching_process} for a schematic representation of this time-dependent branching process). Note that this branching process differs from the widely studied time-dependent branching process \cite{grimmett,age1,age2} because the infectious individual can infect new individuals at any time during his infectious period and not just at the end of its infectious period.

Let the durations of the infectious periods be distributed according to
\begin{equation}
D(t)=\left[\prod_{t'=1}^{t-1}p(t')\right][1-p(t)].
\end{equation} 
Moreover let $\pi(n)$ be the distribution of the avalanche sizes  started by a single infected individual.
The branching process is described by the distribution $\pi(n)$, or equivalently, by its generating function $H_1(x)$ defined as 
\begin{equation}
H_1(x)=\sum_{n=1}^{\infty}\pi(n)x^n.
\label{eq:H1def}
\end{equation}

Assuming that  $t$ is the duration of the infectivity of the seed individual, and  that at each time {$1\leq {t'}< t$} the individual infects $z_{t'}$ other individuals drawn from a Poisson distribution, {\it i.e.} $z_{t'}\sim Poisson(\lambda m(t'))$ the size of the avalanche $n$ generated by the seed individual is given by one plus the sum of the avalanches $n_j^{t'}$ generated by each of the individuals $j$ infected by the seed individual at time ${t'}$.          
Therefore the distribution $\pi(n)$ can be expressed as 
  \bea
\hspace*{-10mm}\pi(n)=\sum_{t=1}^{\infty}D(t)\sum_{\{z_{t'}\}}\sum_{\{n_j^{t'}\}}\left\{\prod_{{t'}=1}^{t-1}\left[P_{t'}(z_{t'})\prod_{j=1}^{z_{t'}}\pi(n_j^{t'})\right]\delta\left(n,\sum_{{t'}=1}^{{t-1}}\sum_{j=1}^{z_{t'}}n_j^{{t'}}+1\right)\right\},
\eea
where $P_t(z)$ is given by 
\bea
P_t(z)=\frac{(\lambda m(t))^z}{z!}e^{-\lambda m(t)}.
\label{eq:Pt}
\eea
The latter recursive equation can be rewritten  by using generating function  $H_1(x)$ as
\bea
H_1(x)=x(F(H_1(x))),\quad \mbox{with}\quad F(x)=\sum_{t=1}^{\infty}D(t)\prod_{t'=1}^{t-1} G_{0,{t'}}(x), \label{eq:H1x}
\eea
where $G_{0,{t}}(x)$ is the generating function of $P_t(z)$
\bea
G_{0,{t'}}(x)=\sum_{z=0}^{\infty}P_{t'}(z)x^{z}=e^{\lambda m(t')(x-1)},
\eea 
where in the last expression we have used the explicit form of   $P_t(z)$ given by Eq. (\ref{eq:Pt}). Therefore it follows that  $F(x)$ is given by
\bea
F(x)=\sum_{t=1}^{\infty} D(t) e^{\lambda M(t)(x-1)},
\eea
where $\lambda M(t)$, indicating the expected total number of primary infected individuals, is given by Eq. (\ref{eq:Mt}).
Summarizing, we conclude that  the self-consistent equation for the generating function $H_1(x)$ can be written as
\bea
H_1(x)=x(F(H_1(x)))=x\sum_{t=1}^{\infty} D(t) e^{\lambda M(t)(H_1(x)-1)}.
\label{eq:H1self}
\eea
\subsection{Relevant kernels}
Let us consider different kernels for both $D(t)$ and $M(t)$. The $D(t)$ kernel  that we will take under consideration are
\begin{itemize}
\item[(1)] {\em The exponential kernel}\\
The exponential kernel is characterized by a $p(t)$ equal to a constant
\begin{equation}
p(t)=a,
\end{equation}
with $0<a<1$.
Therefore we obtain 
\begin{equation}
D(t)=a^{t-1}(1-a).
\label{exp}
\end{equation}
\item[(2)]{\em The power-law kernel}\\
The power-law kernel is characterized by a $p(t)$ given by 
\begin{equation}
p(t)=1-\frac{\alpha-1}{t+\alpha-1},
\end{equation}
with $\alpha>1$ leading to the asymptotic scaling

\bea
D(t)=(\alpha-1)\Gamma(\alpha)\frac{\Gamma(t)}{\Gamma(t+\alpha)}\simeq (\alpha-1) \Gamma(\alpha)t^{-\alpha} \label{eq:power}
\eea
where the last expression indicates the asymptotic scaling valid for  $t\gg 1$.
\end{itemize}
The  $M(t)$ kernels that we will  consider are:
\begin{itemize}
\item[(A)] {\em The linear kernel}\\
The linear $M(t)$ kernel is characterized by a constant $m(t)$,
\bea
m(t)=\bar{m}.
\eea
Therefore we obtain
\bea
M(t)=\sum_{t'=1}^{t-1}m(t')=\bar{m}(t-1)\simeq \bar{m}t,
\eea
where the last expression refers to the asymptotic scaling valid  for $t\gg 1$.
\item[(B)] {\em The power-law decaying kernel}\\
The power-law kernel is characterized by decaying $m(t)$ given by
\bea
m(t)=\tilde{m}\eta\frac{1}{t^{1-\eta}}.
\eea 
Therefore for large time limit, $M(t)$ admits the power-law decay
\bea \label{eq:powerlaw_Mt}
M(t)=\tilde{m}\eta\sum_{t'=1}^{t-1}\frac{1}{t'^{1-\eta}}\simeq \bar{m}{t^\eta},
\eea
where the last expression indicates the asymptotic scaling of $M(t)$ valid for $t\gg 1$.
\end{itemize}
 In the following section we will characterize the critical behaviour of this branching process and its dependence on the different kernels that can be adopted for the functions $D(t)$ and the function $M(t)$.
 
\section{Epidemic threshold of the considered epidemic spreading model}

The time-dependent branching process with containment measures displays finite avalanches whose distribution is fully described by the self-consistent equation for its generating function $H_1(x)$, {\it i.e.} Eq. (\ref{eq:H1self}).\\
Depending on the value of the infectivity $\lambda$ and the expected number $\Avg{M}=\sum_{t=1}^{\infty}D(t)M(t)$ of  primary infections of the seed individual during the entire duration of its infective period we distinguish the three phases of the considered epidemic model.
\begin{itemize}
\item
When $\lambda \avg{M}<1$,  we are in the {\em subcritical phase}.
In this phase all avalanches of the branching process are finite, {\it i.e.} $H_1(1)=1$ and   the expected size of the outbreak started from a single infected individual  is given by:
\begin{equation}
\avg{n}=H'(1)=\frac{1}{1-\lambda \avg{M}}.
\end{equation}
\item
When $\lambda \avg{M}>1$, we are in the {\em supercritical phase}. In this phase   there is a positive probability $S$ that the branching process does not stop, leading to finite avalanches only with probability $H_1(1)=1-S$  where $S\in(0,1]$ is the unique solution of
 \begin{equation}\label{eq:S}
 S = 1 - F(1 - S).
 \end{equation}
 \item
When $\lambda \Avg{M}=1$ we are in the {\em critical phase} characterized by having  $F'(x)=1$, which corresponds to the {\em epidemic threshold} $\lambda_c$ given by 
\begin{equation}\label{eq:criticality}
\lambda_c  =\frac{1}{\avg{M}}
\end{equation}
which is greater than zero as long as $\avg{M}$, is finite.
As $\lambda\to \lambda_c^{\pm}$ the average size of the finite component diverges as 
\bea
\avg{n}\propto \frac{1}{|\lambda-\lambda_c|^{{\gamma},{\gamma}^{\prime}}},
\eea
with ${\gamma}=1$ and ${\gamma}^{\prime}=1$ indicating the critical exponents for $\lambda\to \lambda_c^{-}$ and for $\lambda\to \lambda_c^{+}$ respectively. 
\end{itemize}
Let us now establish the epidemic threshold $\lambda_c$ for the different kernels taken in consideration. In particular we are interested in determining when the epidemic threshold is finite and greater than zero, and when it is zero. In fact a zero epidemic  threshold implies that the epidemic will be always in the supercritical phase, {\it i.e.} for any arbitrarily small value of the infectivity $\lambda$ the epidemic spreads over the population affecting an infinite number of individuals.
Depending on the adopted kernels for $D(t)$ and $M(t)$ the epidemic threshold can be finite or zero:
\begin{itemize}
\item[(1A)] {\em  Exponential $D(t)$ kernel and $M(t)=\bar{m}(t-1)$}\\
Let us  consider the exponential kernel with $D(t)$ given by Eq. (\ref{exp}) and assume $M(t)=\bar{m}(t-1)$.
The expected number of contacts $\avg{M}$ of a random individual is given by 
\bea
\avg{M}=\bar{m}\sum_{t\geq 1}D(t)(t-1)=\bar{m}(1-a)\sum_{t\geq 1}a^{t-1}(t-1)=\bar{m}\frac{a}{1-a}
\eea
The critical threshold $\lambda_c$ is finite for every value of $a\in (0,1)$ with 
\bea
\lambda_c=\frac{1}{\avg{M}}=\frac{1-a}{a\bar{m}}.
\eea
\item[(2A)] {\em Power-law  $D(t)$ kernel, and $M(t)=\bar{m}(t-1)$}\\
Let us  consider the exponential kernel with $D(t)$ given by Eq. (\ref{eq:power}) with $\alpha>1$ and the kernel $M(t)=\bar{m}(t-1)$.
For $\alpha>2$ the expected number of contacts of a random individual is finite and given by
\bea
\avg{M}=\bar{m}\sum_{t\geq 1}(\alpha-1)\Gamma(\alpha)\frac{\Gamma(t)}{\Gamma(t+\alpha)}(t-1)=\bar{m}\frac{1}{\alpha-2}.
\eea
Therefore as long as $\alpha>2$  the epidemic threshold is finite and  given by 
\bea
\lambda_c=({\alpha-2})\frac{1}{\bar{m}}.
\eea
However for  $\alpha\to 2$, $\avg{M}$ diverges and the epidemic threshold $\lambda_c$ vanishes, {\it i.e.}
\bea
\lambda_c=0.
\eea
The epidemic threshold remains zero for all values of $\alpha\in (1,2]$.

\item[(1B)] {\em  Exponential $D(t)$ kernel and power-law decaying $M(t)$ kernel.}\\
Let us consider for $D(t)$ the exponential kernel and for $M(t)$ the power-law kernel with  $\eta\in (0,1)$ (as $\eta=1$ reduces to the constant kernel).
In this case the expected number of primary infections $\avg{M}$ is finite and given by 
\bea
\avg{M}=\sum_{t\geq 1}D(t)M(t)=\frac{\bar{m}}{\eta}\mbox{Li}_{1-\eta}(a).
\eea
Therefore, the  critical threshold $\lambda_c$ is finite and given by
\bea
\lambda_c = \frac{1}{\avg{M}}=\frac{\eta}{\bar{m}\mbox{Li}_{1-\eta}(a)}.
\eea
\item[(2B)] {\em  Power-law  $D(t)$ kernel and power-law decaying $M(t)$ kernel.}\\
Let us consider for $D(t)$ the power-law kernel and for $M(t)$ the power-law kernel with  $\eta\in (0,1)$ (as $\eta=1$ reduces to the constant kernel).
In this case the expected number of primary infections $\avg{M}$ can be expressed as
\bea
\avg{M}=\sum_{t\geq 1} D(t) M(t)=\frac{\bar{m}}{\eta}\Gamma(a)
\sum_{t'\geq 1}\frac{1}{t'^{1-\eta}}\frac{\Gamma(t'+1).}{\Gamma(t'+a)}
\eea
Since asymptotically we have 
\bea
\frac{1}{t'^{1-\eta}}\frac{\Gamma(t'+1)}{\Gamma(t'+a)}\simeq \left(\frac{1}{t'}\right)^{a-\eta},
\eea
 we conclude that for  $\alpha \leq 1+\eta$, $\avg{M}$ diverges and therefore the epidemic threshold vanishes, {\it i.e.} $\lambda_c=0$; and that  for $\alpha>1+\eta$, $\avg{M}$ converges and therefore the epidemic threshold is finite and non-zero $\lambda_c>0$.
\end{itemize}
\section{Critical indices associated to the size of the critical outbreak}
\subsection{Critical exponent $\beta$}

\begin{figure*}
	\includegraphics[width=\columnwidth]{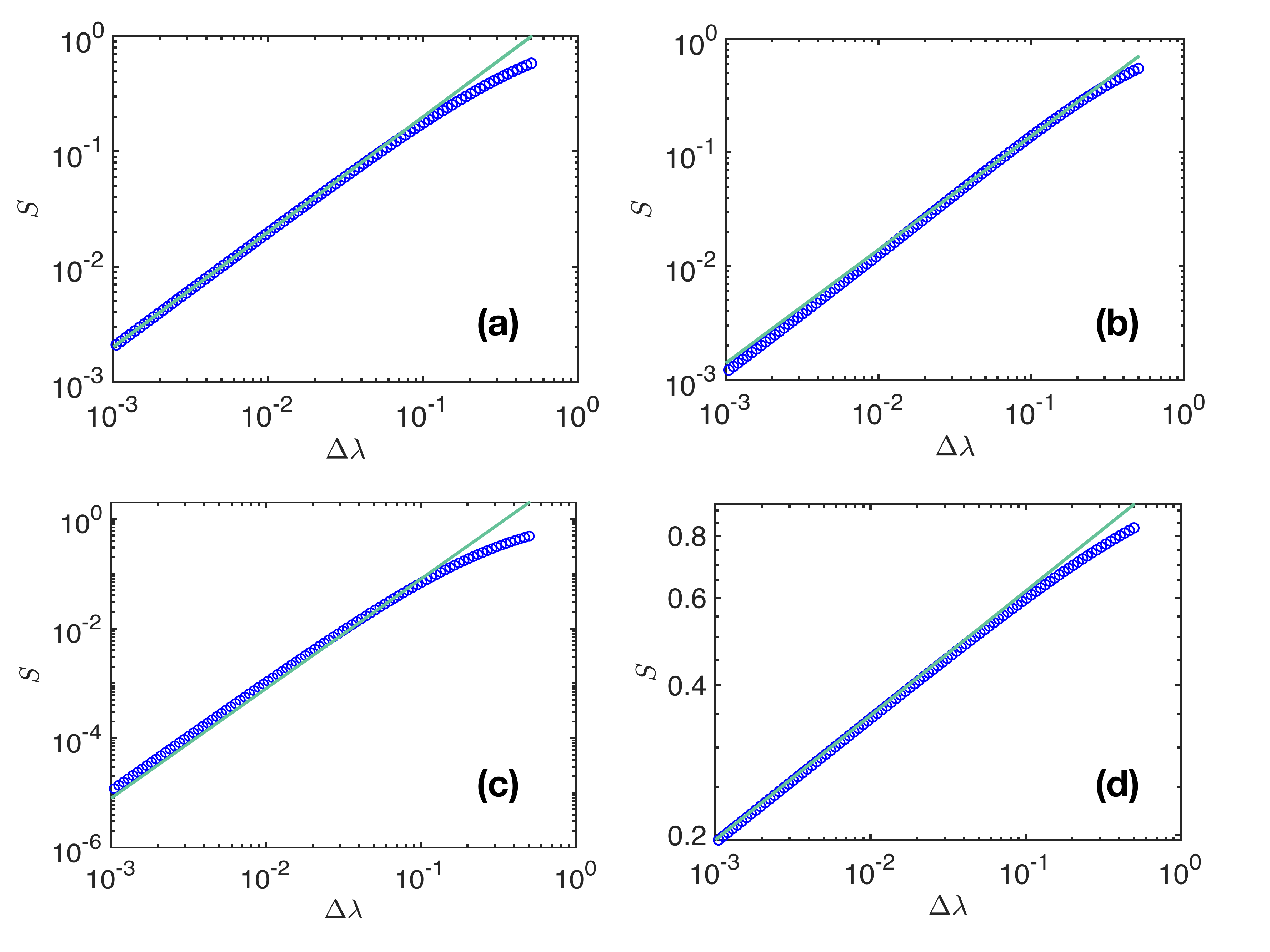} 
	\caption{The probability $S$ that the branching process does not stop is displayed versus the deviation $\Delta \lambda$ of $\lambda$ from the criticality $\lambda_c$. The numerical solutions (blue circles) for exponential the $D(t)$ kernel with $a=0.01$ (a), power-law $D(t)$ kernel with exponent $\alpha=3.5$ (b), $\alpha=2.5$ (c) and $\alpha=1.2$ (d) are obtained by solving Eq.(\ref{eq:S}) numerically and considering always  the $M(t)=\bar{m}(t-1)$ kernel. The predicted asymptotic scaling given by Eq. (\ref{beta_scaling}) using the analytically derived $\beta$ exponents are shown as reference (green lines).
	 }\label{fig:beta}
\end{figure*}

The branching process undergoes a second order phase transition characterized by the order parameter $S=1-H(1)$ indicating the probability of non-extinction of the branching process, with $S$ satisfying Eq. (\ref{eq:S}).
The critical exponent $\beta$ characterizes the scaling of the probability $S$ of observing an infinite avalanche, as a function of $\lambda$ in the critical window $0<\lambda-\lambda_c\ll 1$, in which $S\ll 1$  can be approximated by
\bea
S\simeq A(\lambda-\lambda_c)^{\beta},
\label{beta_scaling}
\eea
with $A$ being a positive constant, {\it i.e.} $A>0$.
This scaling can be predicted to hold in mean-field situations in which $M(t)$ has finite moments, however when some moment diverges the scaling can acquire some logarithmic corrections as we will investigate in the following. Let us predict analytically the scaling of the exponent $\beta$ of the studied epidemic model for the different kernels under consideration.
\begin{itemize}
\item[(1A\&B)] {\em  Exponential $D(t)$ kernel }
When the $D(t)$ kernel is exponential, {\it i.e.}, it is given by Eq. (\ref{exp}), independently on the choice of the kernel for $M(t)$ we are in the mean-field regime where all the moments 
\bea
\avg{M^k}=\sum_{t\geq1}D(t)M^k(t),
\eea are finite. 
In this regime, in order to find the critical exponent $\beta$ we  expand Eq. (\ref{eq:S}) up to the second order in $S$, obtaining
\bea
S\simeq 1-\left[F(1)-F^{\prime}(1)S+\frac{1}{2}F^{\prime\prime}(1)S^2\right],
\eea
where $F(1)=1,F^{\prime}(1)=\lambda \Avg{M}$ and $F^{\prime\prime}(1)=\lambda^2\Avg{M^2}$.
For $0<\lambda-\lambda_c\ll1$ we obtain that $S$ scales according to Eq. (\ref{beta_scaling}) with the mean-field critical exponent $\beta$ given by 
\bea
\beta=1,
\eea
and with $A$ given by 
\begin{equation}
A= 2\frac{\avg{M}^3}{\avg{M^2}}.
\end{equation}

\item[(2A)] {\em Power-law $D(t)$ kernel  and linear $M(t)$ kernel }\\
For the power-law $D(t)$ kernel, the critical exponent $\beta$ can deviate from the mean-field value $\beta=1$ and in general depends on the power-law exponent $\alpha$. Furthermore, for certain values of $\alpha$ the scaling of $S$ in Eq. (\ref{beta_scaling}) develops logarithmic corrections.

Let us consider the linear kernel $M(t)=\bar{m}t$ and the power-law kernel $D(t)$ with power-law exponent $\alpha>1$. The critical index $\beta$ will depend on the value of the power-law exponent $\alpha$.
\begin{itemize}
\item[(i)] For $\alpha > 3$, both $\langle M \rangle$ and $\langle M^2 \rangle$ are convergent, resulting in the mean-field critical exponent $\beta$ given by
\bea
\beta = 1.
\eea
\item[(ii)] For $\alpha=3$ we observe logarithmic corrections to the critical scaling given by Eq.(\ref{beta_scaling}).
Indeed by performing the asymptotic expansion of the  self-consistent equation for $S$ given by Eq. (\ref{eq:S}) for $0<S\ll1 $ we obtain
\bea
S\simeq \lambda \langle M \rangle S -  S^{2}\ln (1/S) \lambda^2 I_0
\eea
where $I_0$ is a constant.
Since, according to Eq. (\ref{eq:criticality}) the epidemic threshold is given by $\lambda_c=1/\Avg{M}$ we obtain that for $0<\lambda-\lambda_c\ll 1$, $S$ follows the scaling
\bea
S\simeq A\frac{(\lambda-\lambda_c)^{\beta}}{\ln [1/(\lambda-\lambda_c)]}
\eea
with $A$ indicating a constant, and $\beta=1$. 
\item[(iii)] For $\alpha \in (2,3)$, the first moment $\Avg{M}$ is convergent, however the second moment $\Avg{M^2}$ is divergent.
We perform the asymptotic expansion of the self consistent equation for $S$  (Eq. (\ref{eq:S})) for $0<S\ll 1$ leading to 
\bea
S &\simeq \lambda \langle M \rangle S -  S^{\alpha-1} \lambda^{\alpha-1} I_1 
\eea
where $I_1$ is a finite constant.
 According to Eq. (\ref{eq:criticality}) we have $\lambda_c \langle M \rangle=1$. Therefore we deduce that $S$ scales follows the critical scaling  given by Eq. (\ref{beta_scaling}) with 
the critical exponent $\beta$ is given by 
\bea\label{eq:beta_divM2}
\beta= \frac{1}{\alpha-2}.
\eea
\item[(iv)] For $\alpha=2$ we observe logarithmic corrections to the critical scaling given by Eq.(\ref{beta_scaling}).
Indeed the asymptotic expansion of Eq. (\ref{eq:S}) for $S\ll1 $ reads,
\bea
  S & \simeq c S \lambda \ln (1/S) I_2
\eea
where $I_2$ is a constant.
By noticing that for $\alpha=2$ the epidemic threshold vanishes, {\it i.e.} $\lambda_c=0$ we deduce that close to criticality, for  $0<\lambda\ll1$ the order parameter $S$ follows the scaling
\bea
S\simeq e^{-A/\lambda}
\eea
where $A$ is a constant.
 \item[(v)] For $\alpha \in (1,2)$, both $\Avg{M}$ and $\Avg{M^2}$ are divergent.
In this case the asymptotic expansion of Eq. (\ref{eq:S}) determining the value of $S$ reads
\bea
  S & \simeq c S^{\alpha-1} \lambda^{\alpha-1} I_3
\eea
where $I_3$ is a finite constant.
Due to the diverging $\langle M \rangle$ the epidemic threshold vanishes, {\it i.e.}  $\lambda_c=0$. Therefore, $S$ scales as Eq. (\ref{beta_scaling})  with critical exponent $\beta$  given by 
\bea\label{eq:beta_divM}
\beta = \frac{\alpha-1}{2-\alpha}.
\eea
\end{itemize}

\item[(2B)] {\em  Power-law $D(t)$ kernel and power-law $M(t)$ kernel }
 
Here we derive the critical exponent $\beta$ for the branching process with   the power-law $M(t)$ kernel 
with $\eta \in (0,1)$ and the power-law $D(t)$ kernel with power-law exponent $\alpha>1$.
Depending on the values of $\eta$ and $\alpha$ we can observe different critical exponents $\beta$. Note that in the limit in which $\eta\to 1$ we recover the critical exponent $\beta$ obtained in case (1B).
\begin{itemize}
\item[(i)] For $(\alpha-1)/\eta>2$  both $\Avg{M}$ and $\Avg{ M^2}$ are convergent,  therefore we can expand $F(x)$ up to the second order. Inserting this expression into the Eq. (\ref{eq:S}) it is immediate to show that  $S$ follows the critical scaling given by Eq. (\ref{beta_scaling}) and that we recover the mean-field critical exponent $\beta=1$.

\item[(ii)] For $(\alpha-1)/\eta=2$ the first moment $\Avg{M}$ is convergent but the second moment $\Avg{M^2}$ is diverging logarithmically. In this case we found logarithmic deviations from the scaling given by Eq. (\ref{beta_scaling}). 
Indeed the asymptotic expansion of Eq.(\ref{eq:S})  for $0<S\ll1$ is given by 
\bea
S \simeq \lambda \langle M \rangle S - \lambda^2 S^2\ln(1/S) I^\prime_0,
\eea
where $I^{\prime}_0$ is a constant.
This asymptotic expansion, together with the expression of the epidemic threshold $\lambda_c=1/\Avg{M}$, leads to the critical scaling of $S$, valid for $0<\lambda-\lambda_c\ll1$ given by 
\bea
S\simeq A\frac{(\lambda-\lambda_c)}{\ln[1/(\lambda-\lambda_c)]}
\eea
where $A$ is a constant.
\item[(iii)] For $(\alpha-1)/\eta \in (1,2)$, the first moment $\Avg{ M}$ is convergent while the second moment $\Avg{M^2}$ diverges.  The epidemic threshold $\lambda_c$ is given by  $\lambda_c=1/\Avg{M}$ and the asymptotic expansion of Eq. (\ref{eq:S}) for $0<S\ll1$ is given by .
\bea
S \simeq \lambda \langle M \rangle S - \lambda^\frac{\alpha-1}{\eta} S^\frac{\alpha-1}{\eta} I^\prime_1,
\eea
where $I^{\prime}_1$ is a constant.
Therefore this asymptotic expansion leads to the critical scaling given by Eq. (\ref{beta_scaling}) with critical exponent
\bea
\beta = \frac{\eta}{\alpha-1-\eta}.
\eea
\item[(iv)] For $(\alpha-1)/\eta=1$ both $\Avg{M}$ and $\Avg{M^2}$ are diverging.
The asymptotic expansion of Eq. (\ref{eq:S}) for $0<S\ll1$ is given by 
\bea
S \simeq   \lambda S\ln (1/S) I^\prime_2,
\eea
where $I^{\prime}_2$ is a constant.
Given that the epidemic threshold in this case is vanishing $\lambda_c=0$ we get that close to criticality, for $0<\lambda\ll1$ $S$ scales like
\bea
S\simeq  e^{-A/\lambda},
\eea
where $A$ is a constant.
\item[(v)] For $(\alpha-1)/\eta \in (0,1)$, both first moment $\Avg{M}$ and second moment $\Avg{M^2}$ are diverging. In this case the epidemic threshold vanishes, {\it i.e.}  $\lambda_c=0.$
The asymptotic expansion of Eq. (\ref{eq:S}) for $0<S\ll1$ is given by 
\bea
S \simeq   \lambda^\frac{\alpha-1}{\eta} S^\frac{\alpha-1}{\eta} I^\prime_3,
\eea
where $I^{\prime}_3$ is a constant.
It follows that in this case, as long as $0<\lambda-\lambda_c\ll1$ the order parameter $S$ follows the critical scaling given by Eq. (\ref{beta_scaling}) with critical exponent $\beta$ given by
\bea
\beta = \frac{\alpha-1}{\eta+1-\alpha}.
\eea
\end{itemize}
\end{itemize}

\subsection{Critical exponents $\tau$ and $\sigma$}
\label{sec:tau}
At criticality the avalanche size distribution $\pi(n)$ follows a power-law scaling with exponent $\tau$ whose value depends on the statistical properties of the $D(t)$ and the $M(t)$ kernels.
Close to criticality the avalanche size distribution $\pi(n)$   acquires a cutoff determined by a scaling function $\Phi(x)$. Specifically for  $\lambda=\lambda_c+\Delta \lambda$,  the avalanche size distribution $\pi(n)$ scales as
\bea
\pi(n)\simeq n^{-\tau} \Phi\left(n (\Delta\lambda)^\sigma \right),
\label{eq:pin}
\eea
where the function $\Phi(x)$ approaches a constant value for $x\to 0$ and decays do zero faster than any power for $x\to\infty$.
In this section, we will derive the critical exponents $\tau$ and $\sigma$ for the different kernels under investigation starting from the self-consistent Eq. (\ref{eq:H1self}) for the generating function $H_1(x)$. We will show that the critical exponents will depend on the choice of the $D(t)$ and the $M(t)$ kernels. However, we notice here that  the scaling relation \cite{cohen2002percolation}
\bea
\sigma(\tau-1)=\beta,
\eea
relating the critical exponents $\sigma, \tau$ to the critical exponent $\beta$
will continue to be satisfied for every choice of the $D(t)$ and $M(t)$ kernels as long as the asymptotic expansion of $F(x)$ for $0<1-x\ll1$ does not have logarithmic corrections.
In order to derive the value of the critical exponent $\tau$ and $\sigma$, determining the scaling of $\pi(n)$ according to Eq. (\ref{eq:pin}), we first observe this scaling implies that the generating function $H_1(x)$ defined as Eq.(\ref{eq:H1def}) for $0<1-x\ll 1$ scales as  
\bea
H_1(x) \simeq 1-(1-x)^{\tau-1}h\left(\frac{1-x}{(\Delta \lambda)^\sigma}\right).
\label{eq:H1_scaling}
\eea
where $h(x)$ is a scaling function \cite{marsili1,marsili_bianconi}.
By inserting this scaling relation into the self consistent equation for $H_1(x)$ (Eq.(\ref{eq:H1self})) which we rewrite here for convenience,
\bea
H_1(x)=xF(H_1(x)),
\label{eq:H1self2}
\eea 
we will the critical exponents $\tau$ and $\sigma$ for all the kernels under consideration.

%\end{document}

\begin{itemize}
\item[(1A\&B)] {\em Exponential $D(t)$ kernel}\\
With an exponential $D(t)$ kernel, all the moments of $M(t)$ are finite. Therefore we are in the mean-field regime, which is independent on the choice of $M(t)$ kernel.
We consider the self-consistent equation for $H_1(x)$ given by Eq.(\ref{eq:H1self2}) where we substitute the scaling of $H_1(x)$ for $0<1-x\ll 1$ given by Eq.(\ref{eq:H1_scaling}). In the case in which $0<1-x\ll1$ we have $0<1-H_1(x)\ll1$, therefore in Eq. (\ref{eq:H1self2}) we can substitute $F(w)$ with this Taylor expansion around $w=1$ truncated at the second order. 
By putting $1-x=z(\Delta \lambda)^{\sigma}$  we get
\bea
F(H_1(x))&\simeq& 1-\lambda\avg{M} z^{\tau-1}(\Delta \lambda)^{\sigma(\tau-1)}h(z)\nonumber \\
&&+\frac{1}{2}\lambda^2\Avg{M^2} z^{2(\tau-1)}(\Delta \lambda)^{2\sigma(\tau-1)}h^2(z).
\eea
Inserting this expression into Eq. (\ref{eq:H1self2}) and using the explicit expression of the epidemic threshold $\lambda_c=1/\Avg{M}$, we get for $0<\lambda-\lambda_c\ll1$ 
\bea
\hspace*{-8mm}\avg{M}z^{\tau-1}(\Delta \lambda)^{\sigma(\tau-1)+1}h(z)+z(\Delta \lambda)^{\sigma}-cz^{2(\tau-1)}(\Delta \lambda)^{2\sigma(\tau-1)}h^2(z)=0,
\eea
where $c$ is a constant given by 
$c=\lambda_c^2\Avg{M^2}/2$.
Imposing that all the terms in the above expansion are of the same order, {\it i.e.} putting 
\bea
\sigma(\tau-1)+1=\sigma=2\sigma(\tau-1),
\eea
we get the mean-field critical exponents
\bea
\tau=3/2,\quad \sigma=2.
\eea

\item [(2A)] {\em Power-law $D(t)$ kernel and linear $M(t)$ kernel}\\
When the  $D(t)$ kernel is power-law with power-law exponent $\alpha>1$, the exponents $\tau$ and $\sigma$  depend on the value of $\alpha$ and can deviate from the mean-field values.
In the following we evaluate the exponents $\tau$ and $\sigma$ for values of the exponent $\alpha$ that lead to an expansion of $F(x)$ for $0<1-x\ll 1$ that does not have logarithmic corrections. 
\begin{itemize}
\item[(i)] For $\alpha > 3$, both $\avg{M}$ and $\avg{M^2}$ are finite. By expanding $F(x)$ for $0<1-x\ll 1$ up to the second order   we can reproduce the calculation performed for the exponential $D(t)$ kernel. Therefore we recover the mean-field critical exponents
\bea
\tau=3/2,\quad \sigma=2.
\eea

\item[(ii)] For $\alpha \in (2,3)$, $\langle M \rangle$ is convergent and $\langle M^2 \rangle$ is divergent, while the epidemic threshold is finite and given by  $\lambda_c =1/\Avg{M}$.
We consider the asymptotic expansion of $F(w)$ for $w=H_1(x)$ and $0<1-x\ll 1$ given by  
\bea
F(H_1(x)) & \simeq 1 - z(\Delta \lambda)^{\sigma({\tau-1})} h(z) \lambda \langle M  \rangle \nonumber \\
& +   \lambda^{\alpha-1} \left[ \left(z(\Delta \lambda)^\sigma\right)^{\tau-1} h(z)\right]^{\alpha-1} I_1,
\eea
where $I_1$ is a constant.
By inserting this expression in the self consistent formula for $H_1(x)$ given by Eq.(\ref{eq:H1self2}) we get the leading terms
\bea
\hspace*{-20mm}\avg{M}z^{\tau-1}(\Delta \lambda)^{\sigma(\tau-1)+1}h(z)+z(\Delta \lambda)^{\sigma}-\lambda_c^{\alpha-1}I_1\left[\left(z(\Delta \lambda)^{\sigma}\right)^{(\tau-1)}h(z)\right]^{\alpha-1}=0,\nonumber 
\eea
Imposing  that all the terms in the above equation are of the same order,
\bea
\sigma(\tau-1)+1 =\sigma= \sigma(\tau-1)(\alpha-1),
\eea
 we obtain the critical exponents
\bea
\tau=\frac{\alpha}{\alpha-1},\quad \sigma=\frac{\alpha-1}{\alpha-2}.
\eea
\item[(iii)] For $\alpha \in (1,2)$, both $\Avg{M}$ and $\Avg{M^2}$ are  divergent, while the epidemic threshold is vanishing $\lambda_c=0$.
We proceed by considering the asymptotic expansion of $F(w)$ for $w=H_1(x)$ with $0<1-x\ll1$, with $H_1(x)$ scaling according to Eq. (\ref{eq:H1_scaling}), getting
\bea
F(H_1(x)) \simeq 1- (\Delta \lambda)^{\alpha-1} \left[ \left(z(\Delta \lambda)^\sigma\right)^{\tau-1} h(z)\right]^{\alpha-1} I_3
\eea 
where $I_3$ is a constant.
By inserting this expression in the self consistent formula for $H_1(x)$ given by Eq.(\ref{eq:H1self2}) we get the leading terms
\bea
\hspace*{-20mm}\avg{M}z^{\tau-1}(\Delta \lambda)^{\sigma(\tau-1)}h(z)+z(\Delta \lambda)^{\sigma}-(\Delta \lambda)^{\alpha-1} \left[ \left(z(\Delta \lambda)^\sigma\right)^{\tau-1} h(z)\right]^{\alpha-1} I_3=0.\nonumber
\eea
By  imposing that all the terms in the above equation are of the same order,
\bea
\sigma=\sigma(\tau-1)=\sigma(\tau-1)(\alpha-1)+\alpha-1,
\eea
 we obtain the critical exponents
\bea
\tau=2,\quad \sigma=\frac{\alpha-1}{2-\alpha}.
\eea
\end{itemize}

\item [(2B)] {\em Power-law $D(t)$ kernel and power-law $M(t)$ kernel}
\begin{itemize}

\item[(i)] For $(\alpha-1)/\eta>2$ , both $\avg{M}$ and $\avg{M^2}$ are convergent. Thus we recover the mean-field exponents
\bea
\tau=3/2, \quad \sigma=2.
\eea
\item[(ii)] For $(\alpha -1)/\eta \in (1,2)$, $\Avg{M}$ is convergent, $\Avg{M^2}$ is divergent and the epidemic threshold  is finite and given by  $\lambda_c= 1/\Avg{M}$.
We consider the asymptotic expansion of $F(w)$ for $w=H_1(x)$ and $0<1-x\ll1$: 
\bea
F(H_1(x)) & \simeq &1 - z^{\tau-1}(\Delta \lambda)^{\sigma({\tau-1})} h(z) \lambda \Avg{M}\nonumber \\
& &+  \lambda^\frac{\alpha-1}{\eta} \left[ \left(x(\Delta \lambda)^\sigma\right)^{\tau-1} h(z)\right]^\frac{\alpha-1}{\eta}I^{\prime}_1,
\eea
where $I^{\prime}_1$ is a constant.
By inserting this expansion in the self-consistent Eq. (\ref{eq:H1self2})  we find that the leading terms are given by 
\bea
\hspace*{-25mm}\avg{M}z^{\tau-1}(\Delta \lambda)^{\sigma(\tau-1)+1}h(z)+z(\Delta \lambda)^{\sigma}- \lambda_c^\frac{\alpha-1}{\eta} \left[ \left(x(\Delta \lambda)^\sigma\right)^{\tau-1} h(z)\right]^\frac{\alpha-1}{\eta}I^{\prime}_1=0,
\eea
By imposing that all these terms are of the same order, {\it i.e.} by imposing
\bea
\sigma(\tau-1)+1 =\sigma= \frac{\sigma(\tau-1)(\alpha-1)}{\eta},
\eea
 we obtain the critical exponents
\bea
\tau=\frac{\eta+\alpha-1}{\alpha-1},\quad \sigma=\frac{\alpha-1}{\alpha-1-\eta}.
\eea

\item[(iii)] For $(\alpha-1)/\eta \in (0,1)$, both $\Avg{M}$ and $\Avg{M^2}$ are divergent, and the epidemic threshold vanishes, {\it i.e.} $\lambda_c=0$.
By proceeding like the in three previous cases we consider the asymptotic expansion of $F(w)$ for $w=H_1(x)$ and $0<1-x\ll1$, given by 
\bea
F(H_1(x)) \simeq 1 - (\Delta \lambda)^{\frac{\alpha-1}{\eta}}\left[ (z(\Delta \lambda)^\sigma)^{\tau-1} h(z)\right]^{\frac{\alpha-1}{\eta}}I^\prime_3,
\eea
where $I^{\prime}_3$ is a constant.
By substituting this asymptotic expansion in the self consistent equation for $H_1(x)$ we get to leading order,
\bea
\hspace*{-20mm}\avg{M}z^{\tau-1}(\Delta \lambda)^{\sigma(\tau-1)}h(z)+z(\Delta \lambda)^{\sigma}-(\Delta \lambda)^{\frac{\alpha-1}{\eta}} \left[ (z(\Delta \lambda)^\sigma)^{\tau-1} h(z)\right]^{\frac{\alpha-1}{\eta}}I^\prime_3=0.\nonumber
\eea
Imposing that all the terms of the above equation are of the same order,  by putting
\bea
  \sigma = \sigma(\tau-1) = \frac{\sigma(\tau-1)(\alpha-1)+\alpha-1}{\eta},
\eea
we obtain the critical exponents
\bea
\tau=2, \quad \sigma=\frac{\alpha-1}{\eta-\alpha+1}.
\eea
\end{itemize}
\end{itemize}

\begin{figure*}
	\includegraphics[width=1.05\columnwidth]{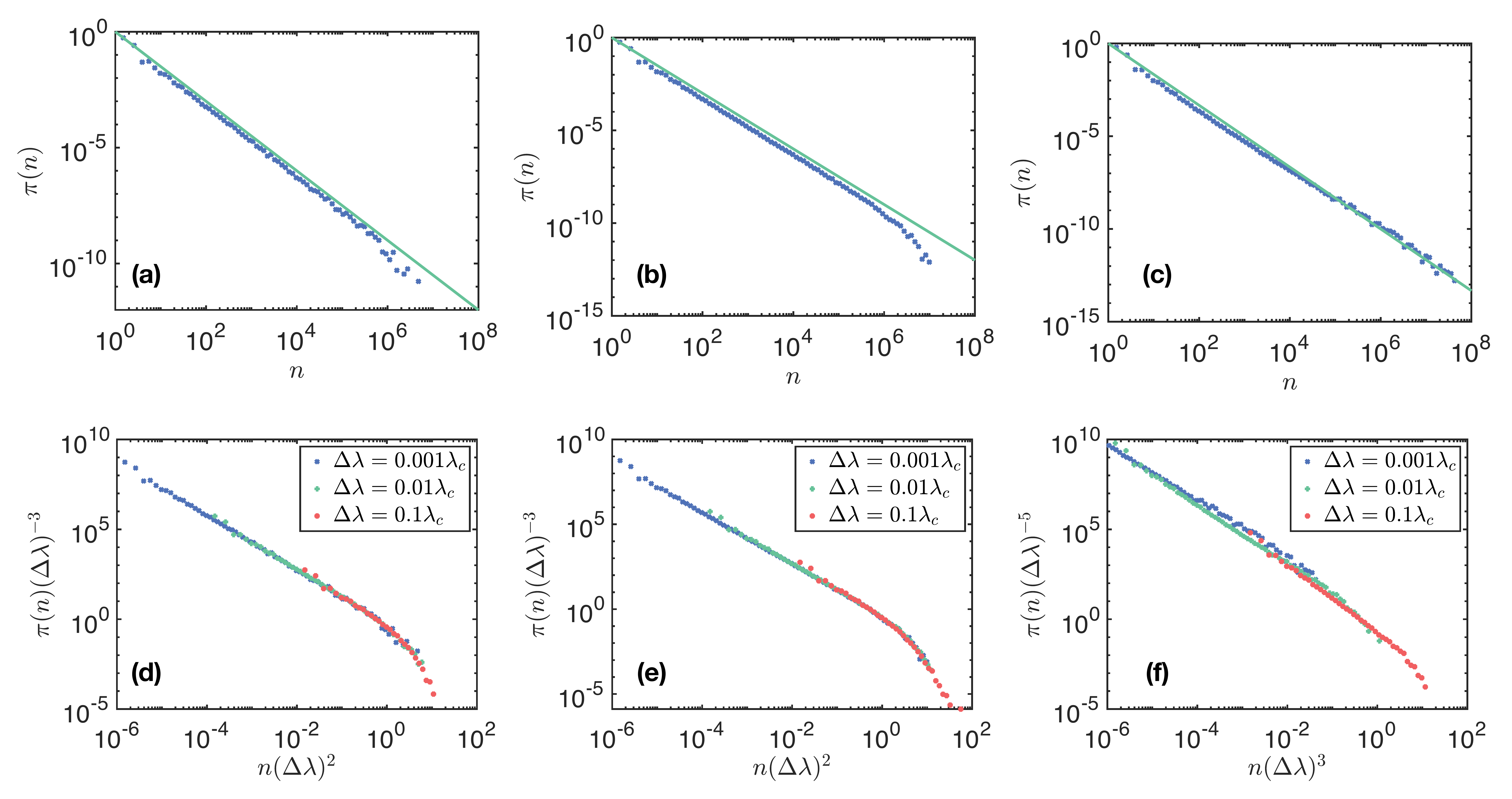} 
	\caption{Monte Carlo simulations of the critical branching processes with exponential and power-law $D(t)$ kernels. Panels (a), (b) and (c) show the critical distribution of avalanche size $\pi(n)$ corresponding to  different $D(t)$ kernels and panel (d), (e) and (f) show the  data collapse for distributions obtained away from criticality for the same $D(t)$ kernels. Panels (a), (d): exponential kernel with $a=0.01$, panels (b), (e) power-law kernel with $\alpha=3.5$, panels (c),(f) power-law kernel with $\alpha=2.5$. The distributions are obtained from simulations of $10^5$  realizations of the branching process with a linear  $M(t)=t$  kernel.
	 }\label{fig:tau}
\end{figure*}

%\end{document}
\section{Distribution of the temporal duration of avalanches}

In the previous section we have shown how the distribution of critical avalanche size depends on the kernel of the considered branching process modelling epidemics spreading with time-dependent containment measures.
Here we show that instead the distribution of the avalanche duration is determined by critical exponents that are independent of the choice of the kernels under consideration.
Let us define $y$ as the critical exponent characterizing asymptotic scaling of the  distribution $P(T)$ of the duration $T$ of critical avalanches 
\bea
P(T)\simeq C^{\prime} T^{-y}.
\eea
for $T\gg 1$  where $C^{\prime}$ is a constant. 
The cumulative distribution of $P(T)$ denoted by $\hat{P}(T)$ indicates the probability that the avalanche has not stopped at time $T$, and scales  for $T\gg 1$ as 
\bea
\hat{P}(T)\simeq C T^{-y+1}.
\label{eq:hPTscaling}
\eea where  $C$ is a  constant.  
We note that a critical avalanche started from a single initial seed is extinct at time $T$ if each subavalanche generated by any of the offspring of the seed node is also extinct. Therefore, it is immediate to show that $\hat{P}(T)$ satisfies 
\bea
1-\hat{P}(T)&=&\sum_{t\geq 1}D(t)\prod_{t'=1}^{t-1}\Avg{\left[1-\hat{P}(T-t')\right]^{z_{t'}}}_{z_{t'}}\nonumber \\
&&=\sum_{t\geq 1}D(t) \exp\left[-\lambda \sum_{t'=1}^{t-1}m(t')\hat{P}(T-t')\right],
\label{eq:PTself}
\eea
where in the last expression we consider average over the Poisson distribution for $z_{t'}$.
In order to determine the exponent $y$, we insert the critical scaling for $\hat{P}(T)$ given by Eq. (\ref{eq:hPTscaling}) into Eq. (\ref{eq:PTself}) and check that this equation is satisfied only for $y=2$ (which is the mean-field exponent) independently of the choice of the $D(t)$ and $M(t)$ kernels.
%\end{document}
To this end,  let us take $m(t)=\bar{m}$ corresponding to  the linear $M(t)$ kernel, and let us consider a generic $D(t)$ kernel.
By inserting the scaling function for $\hat{P}(T)$ given by Eq. (\ref{eq:hPTscaling}) with $y=2$ into the left hand side of the self-consistent Eq. (\ref{eq:PTself})  we get, at the critical point $\lambda=\lambda_c,$
\bea
\hspace{-8mm}\sum_{t\geq 1}D(t) \exp\left[-\lambda_c \sum_{t'=1}^{t}m(t')\hat{P}_c(T-t')\right]\nonumber\\
\simeq \sum_{t\geq 1}D(t) \exp\left[-\lambda_c C \left(\phi^{(0)}(1-T)-\psi^{(0)}(1-T+t)\right)\right],\label{eq:Ptlinear}
\eea
where $\psi^{(0)}(x)$ is the $0$-th PolyGamma function.
We consider the expansion  for $T\gg t$, getting
\bea
\phi^{(0)}(1-T)-\psi^{(0)}(1-T+t)=\frac{t}{T}+O(1/T^2).
\eea
Inserting this expansion in Eq. (\ref{eq:Ptlinear}) we obtain  to leading terms
\bea
\sum_{t\geq 1}D(t) \exp\left[-\lambda_c \sum_{t'=1}^{t}m(t')\hat{P}_c(T-t')\right]=\sum_{t\geq 1}D(t)\exp\left[-\lambda_c C \bar{m}\frac{t}{T}\right]\nonumber \\\simeq 1-\frac{C}{T},
\eea 
where in the last expression we have first expanded for $T\gg1 $ and then we have used $\lambda_c\Avg{M}=1$.
Therefore with this derivation we get   that Eq. (\ref{eq:PTself}) is identically satisfied at criticality with the choice of $\hat{P}(T)$ given by Eq. (\ref{eq:hPTscaling}) as long as  $y=2$.
By considering the power-law $M(t)$ kernel it can be shown that the critical exponent $y=2$ is not modified.  
In fact, taking $m(t)=(\bar{m}{\eta})t^{\eta-1}$ with $\eta\in (0,1)$  we can evaluate  the left hand side of the self-consistent Eq. (\ref{eq:PTself})  
for $T\gg 1$ using continuous approximation to obtain:
\bea
\sum_{t\geq 1}D(t) \exp\left[-\lambda_c \sum_{t'=1}^{t}m(t')\hat{P}_c(T-t')\right]\nonumber \\
\simeq \sum_{t\geq 1}D(t) \exp\left[-\lambda_c \int_0^t \frac{C\bar{m}\eta (t')^{\eta-1}}{T-\tau'} dt' \right]\nonumber \\
=\sum_{t\geq 1} D(t) \exp\left[ -\lambda_c C\bar{m}\eta T^{\eta-1} \mathcal{B}_{t/T}(\eta, 0)\right]\label{eq:beta_function} 
\eea
%\end{document}
where $\mathcal{B}_{t/T}(\eta, 0)$ is the incomplete Beta function.\\
By further considering the expansion of  the Beta function  for $T\gg 1$, given by $\mathcal{B}_{x}(\eta,0)\simeq x^{\eta}$  we get
\bea
\sum_{t\geq} D(t) \exp\left[ -\lambda_c \bar{m}\eta C T^{\eta-1} \mathcal{B}_{t/T}(\eta, 0)\right] \simeq\sum_{t\geq} D(t) \exp\left[ -\lambda_c \bar{m}\eta C\frac{t^{\eta}}{T}\right]\nonumber \\
\simeq \sum_{t\geq} D(t)  \left(1- \lambda_c{C\bar{m}\eta}\frac{t^\eta}{T}\right)= 1-\frac{{C}}{T}
\eea
where we have used $\lambda_c\Avg{M}=1$ with $\avg{M}$ given, in the continuous approximation,  by 
\bea
\Avg{M}=\bar{m}\eta \Avg{t^{\eta}}.
\eea 
Therefore this derivation shows that also for the power-law $M(t)$ kernel  we get that Eq.~(\ref{eq:PTself}) is identically satisfied at criticality provided $y=2$.

\begin{figure*}
	\includegraphics[width=1.\columnwidth]{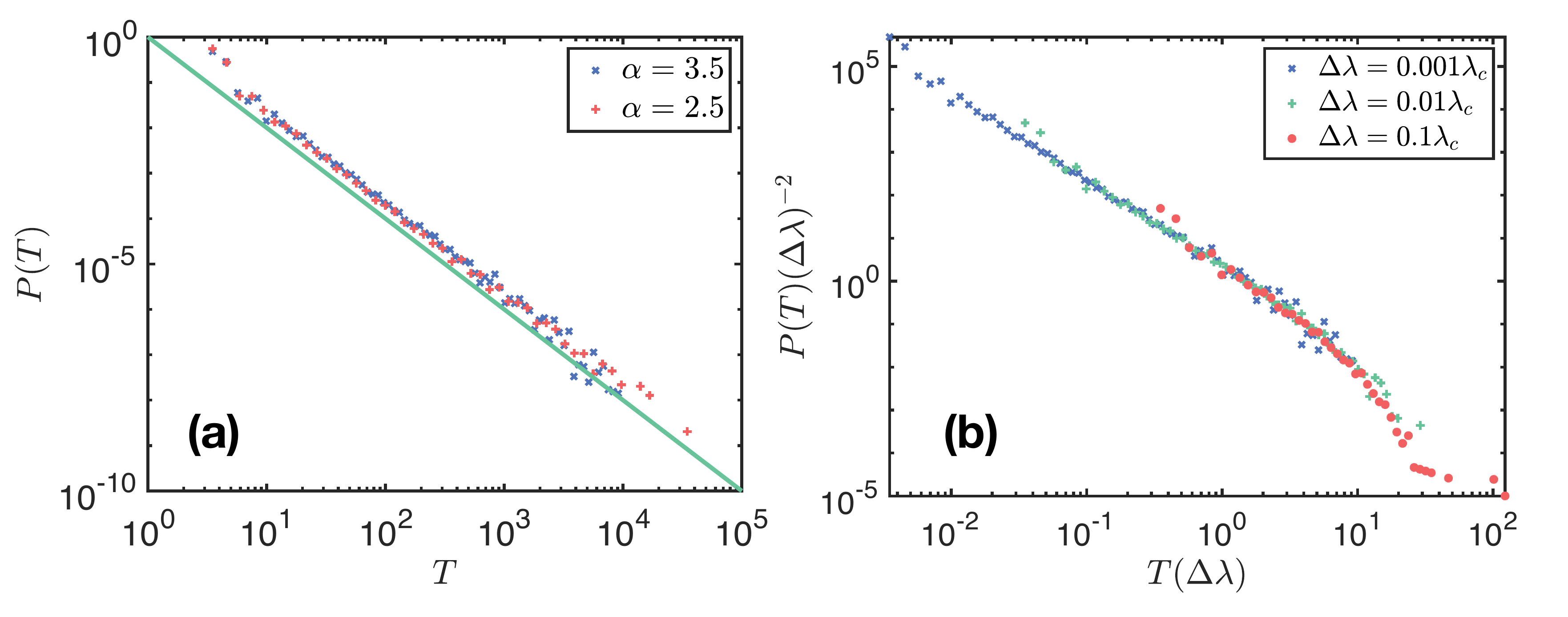} 
	\caption{Monte Carlo simulations of time-duration distribution of critical branching processes with power-law $D(t)$ kernel with $\alpha=3.5$ and $\alpha=2.5$. The distributions of time-duration of avalanches (panel (a)) and the data collapse (panel (b)) with $\alpha=3.5$ are shown. The distributions are obtained from $10^5$ samples of critical branching processes and $M(t)=t$ is considered in the simulations. The distributions with different $D(t)$ kernel give the same critical exponents.
	 }\label{fig:time_distribution}
\end{figure*}

In the sublinear regime, for $0<\Delta \lambda=\lambda_c-\lambda\ll1 $, we can proceed in a similar manner as for the standard branching process \cite{lauritsen1996self,zapperi1995self} and show that the power-law scaling of $P(T)$ is modulated by a function of $T(\Delta \lambda)^{\epsilon}$ with $\epsilon=1$ leading to the scaling 
\bea
P(T)\simeq \frac{1}{T^2}\Psi(T(\Delta \lambda)),
\eea
where $\Psi(x)$ converges to a constant for $c\to 0$ and decays exponentially  for $x\to \infty$.
 These predictions agree perfectly with the Monte Carlo simulations (See Figure.\ref{fig:time_distribution}).

\section{Dynamics of the critical branching process}

\begin{figure*}\center
	\includegraphics[width=0.6\textwidth]{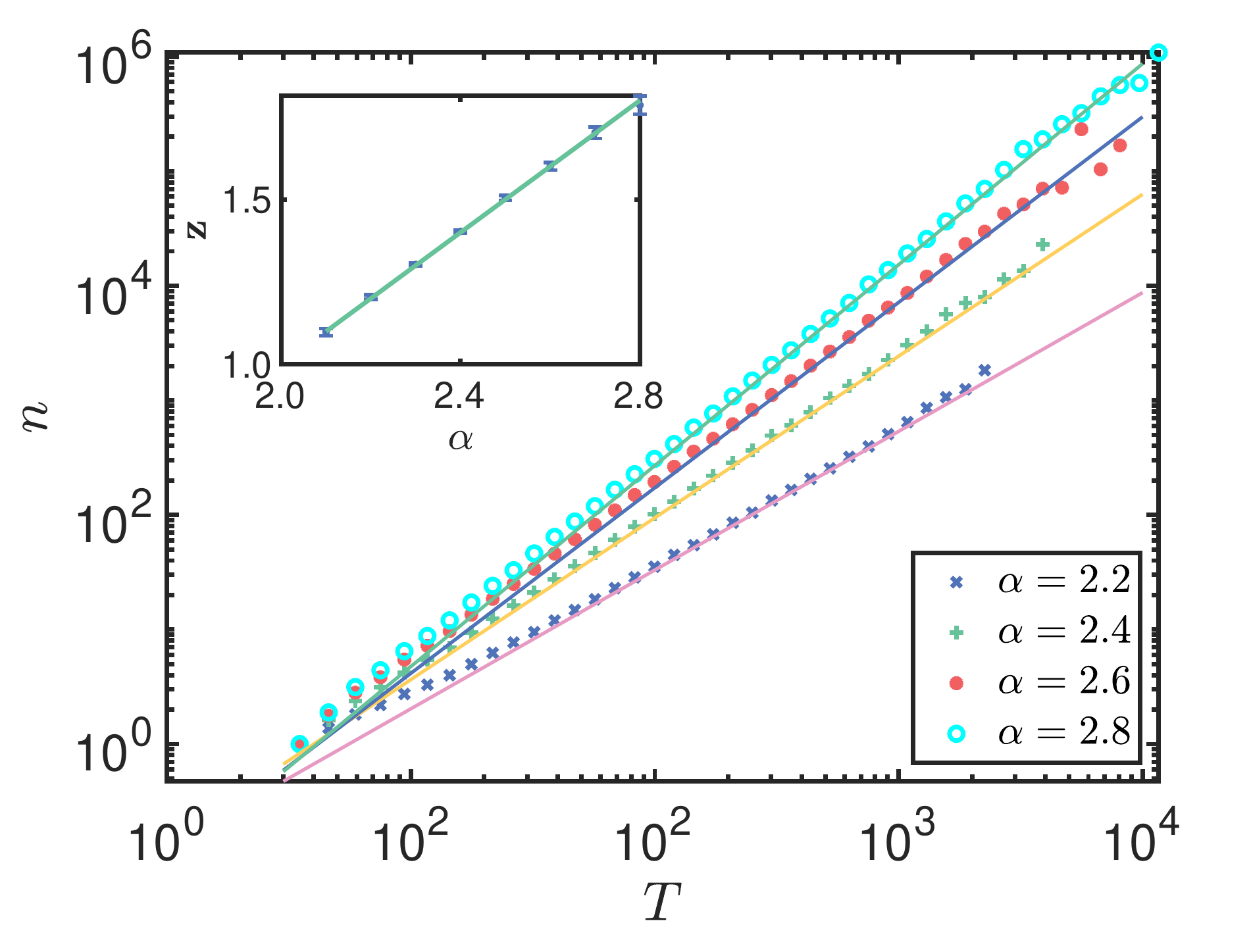} 
	\caption{The size of the avalanches $n$ is shown versus the time-duration of the avalanches $T$ for the  critical branching process with $D(t)$ power-law kernel with different power-law exponents $\alpha$. The $M(t)$ kernel is linear. Symbols indicate numerical simulations, solid lines indicate power-law fit to the data. Inner panel: The fitted power-law exponent $z$ (blue dots) is shown versus $\alpha$ and compared with the  theoretical expectation $z=\alpha-1$ (green line).}\label{fig:RT_dependence}
\end{figure*}

\begin{figure*}\center
	\includegraphics[width=1.05\textwidth]{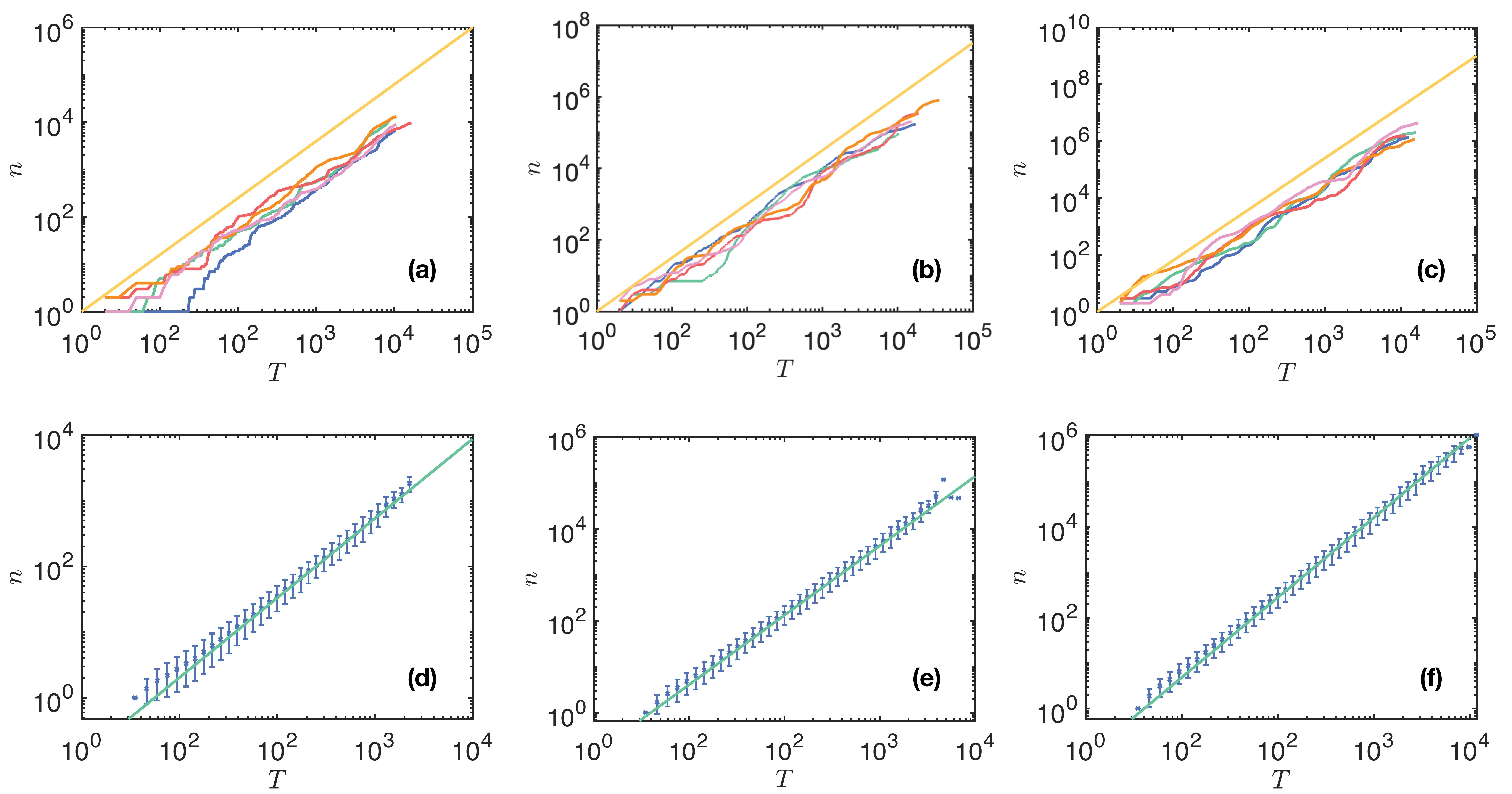} 
	\caption{The size of the avalanches $n$ is shown versus the time-duration of the avalanches $T$ for five single instance realizations of the  the  critical branching process with $D(t)$ power-law kernel and power-law exponents  $\alpha=2.2$ (panel (a)), $\alpha=2.5$ (panel (b)) and $\alpha=2.8$ (panel (c)). In panels (a-c) the theoretical expectation is  shown as a yellow line. In panel (d)-(f) we show the size of the avalanches $n$ versus the time-duration of the avalanches $T$ averaged over $10^6$  realizations of the critical branching process with $D(t)$ power-law kernel and power-law exponents  $\alpha=2.2$ (panel (c)), $\alpha=2.5$ (panel (d)) and $\alpha=2.8$ (panel (e)) explicitly indicating the errorbars. }\label{fig:single_instance_errorbar}
\end{figure*}

At criticality, the avalanche size $n$ is related to the duration of the avalanche by a power-law scaling determined by the   
critical dynamic exponents by $z$, {\it i.e., }
\bea
n \propto T^{z}.
\label{eq:nz}
\eea
This power-law dependence of $n$ with $T$ is only observed exactly at criticality, for $\lambda=\lambda_c$ while in the supercritical phase we have an exponential growth of the individual of an avalanche in time.
The dynamical exponent $z$ can be easily found once the exponents $\tau$ and $y$, determining the critical scaling of $\pi(n)$ and $P(T)$, are known.
In fact $z$ can be found by imposing that at criticality, {\it i.e.} for $\lambda=\lambda_c,$ 
\bea
P(T)dT=\pi(n)dn,
\eea
where $n$ scales with $T$ according to Eq.(\ref{eq:nz}), $\pi(n)\propto n^{-\tau}$ and $P(T)\propto T^{-y}$.
In this way, using the fact that $y=2$, it  is straightforward to show that the critical exponent $z$ is given by 
\bea
 z= \frac{1}{\tau-1}.
\eea
It follows that $z$ depends of the choices of the $D(t)$ and $M(t)$ kernels.
\begin{itemize}
\item[1A \& 1B]{\em Exponential $D(t)$ kernel.}
In the case of the exponential $D(t)$ kernel,   we recover the  mean-field exponents 
\bea
\tau=3/2, \quad z=2,
\eea 
both for the linear and the power-law $M(t)$ kernel.
\item[2A \& 2B]{\em Power-law $D(t)$ kernel}.
In the case of  power-law $D(t)$ kernel the dynamical exponent $z$ ranges between one and two, {\it i.e.} $z\in [1,2].$
Let us treat the case of the linear $M(t)\propto t$ kernel and the power-law $M(t)\propto t^{\eta}$ together by taking $\eta\in (0,1]$ where for $\eta=1$ we recover the linear kernel.
When neglecting the  values of $\alpha$ in which the expansion of $F(x)$ around $x=1$ has logarithmic corrections, and considering the values of $\tau$ derived in Sec. $\ref{sec:tau}$, we see that the dynamical exponent $z$ changes as a function of $\alpha$ and $\eta$ in the following way.
\begin{itemize}
\item[(i)]  For $(\alpha-1)/\eta>2$ we recover the mean-field exponents 
\bea
\tau=3/2,\quad z=2.
\eea
\item[(ii)] For $(\alpha-1)/\eta\in (1,2)$ we obtain 
\bea
\tau=1+\frac{\eta}{\alpha-1},\quad z=\frac{\alpha-1}{\eta}.
\eea
It follows then that in particular, for  the linear kernel, {\it i.e.} for $\eta=1$ we obtain $z=\alpha-1$ which agrees  with the numerical simulations {(see Figure \ref{fig:RT_dependence} for comparison of  the theoretical results with  the average over simulations of the critical branching process and Figure \ref{fig:single_instance_errorbar} for comparison of the theoretical results with simulations of single instances of the branching process).}
\item[(iii)] For $(\alpha-1)/\eta\in (0,1)$ we obtain 
\bea
\tau=2,\quad z=1.
\eea
\end{itemize}
\end{itemize}
As mentioned before these dynamical critical exponents agree with extensive Monte Carlo simulations, and display values that can be only obtained by taking into consideration the stochastic effects of the dynamics that play a crucial role of criticality. As a consequence it is possible to observe that the critical exponent $z$ derived here deviates from the corresponding dynamical exponent that can be derived from the deterministic dynamics \cite{bianconi2020epidemics}.

\section{Conclusions}

In this work we have studied a stochastic epidemic model with containment measures  in which each infected individual is infectious for a time $t$ with a given distribution $D(t)$. Additionally, during the infectious period an individual can infect a constant, or time-varying number of individuals resulting in a total number of secondary infections $M(t)$ that either increases linearly or sublinearly with time.
We have shown that depending of the choice of the $D(t)$ and $M(t)$ kernels, the critical behaviour of the branching process that captures this epidemic spreading model changes. In particular the critical index $\tau$ that characterise the distribution  of avalanche sizes depends on the choice of the kernels $D(t)$ and $M(t)$ and ranges in the interval between $3/2$ and $2$, {\it i.e.} $z\in [3/2,2]$.
However, the critical exponent determining the avalanche duration appears to be universal and independent on the choice of the  $D(t)$ and $M(t)$ kernels.
Most relevantly, the study of this model allows us to derive  the expression for the dynamical critical exponent $z$ that determines the power-law growth of the number of infected individuals $n$  and the avalanche duration of critical avalanches $T$, {\it i.e.} $n\propto T^z$.  Interestingly, this critical exponent can be related to empirical observations on COVID data that starting from the work of Ziff and Ziff \cite{ziff2020fractal} have detected power-law increases of the number of cases in time  \cite{maier2020effective,nekovee2020understanding,brandenburg2020piecewise}

We recover the classic results for the dynamical  exponent $z=2$ in the standard branching process, and we predict that containment measures that have the effect of modulating the $D(t)$ and the $M(t)$ kernels can have the effect to modify the value of $z$ allowing $z\in [1,2].$
These theoretical results show that stochastic effects are important when determining the dynamical exponent $z$. Indeed, the exponent found in this paper improves on the deterministic treatment proposed in \cite{bianconi2020epidemics}.
More importantly, the result presented in this work shows that the dynamical critical exponent $z$ can be modulated by time-dependent containment measures in the range $z\in [1,2]$ which is consistent with some empirical observations made during few periods of strong mitigation  of the COVID-19 pandemic observed in the last two years.
We note however that this range does not include the value originally found by Ziff and Ziff in the first work \cite{ziff2020fractal} in which a power-law growth with exponent larger than two of the number of cases in time was reported.
This implies that although containment measures that have the effect of modulating the $D(t)$ and $M(t)$ kernels can tune the value of the critical exponent $z$, other mechanisms including for instance the role of a  (hyperbolic) hierarchical, and nested spatial distribution of the spreading process might be also play a role in determining the actual value of $z$ in real epidemics.
\section*{References}
\bibliographystyle{unsrt}
\bibliography{ref}

\end{document}